\documentclass{aa}

\usepackage{graphicx,times}

\begin{document}

\title{Cassiopeia A: on the origin of the hard X-ray continuum
and the implication of the observed \ion{O}{viii} Ly-$\alpha$/Ly-$\beta$
distribution}

\author{J.A.M. Bleeker\inst{1}
   \and
    R. Willingale\inst{2}
   \and
   K. van der Heyden\inst{1}
    \and
    K. Dennerl\inst{3}
   \and
   J.S. Kaastra\inst{1}
   \and
     B. Aschenbach\inst{3}
   \and
   J. Vink\inst{4,5,6}}


\institute{SRON National Institute for Space Research, Sorbonnelaan 2,
3584 CA Utrecht, The Netherlands
\and
Department of Physics and Astronomy, University of Leicester, University Road,
Leicester LE1 7RH
\and
Max-Planck-Institut f\"ur extraterrestrische Physik, Giessenbachstra{\ss}e,
D-85740 Garching, Germany
\and
Astrophysikalisches Institut Potsdam,
An der Sternwarte 16, D-14482, Potsdam, Germany
\and
Columbia Astrophysics Laboratory, Columbia University,
550 West 120th Street, New York, NY 10027, USA
\and
Chandra Fellow
}

\mail{J.A.M.Bleeker@sron.nl}

\titlerunning{Cas A}
\authorrunning {J.A.M. Bleeker et al.}

\date{Received ; accepted }

\abstract{
We present the first results on the hard X-ray continuum image (up to 15~keV) of
the supernova remnant Cas~A measured with the EPIC cameras onboard XMM-Newton.
The data indicate that the hard X-ray tail, observed previously, that extends to
energies above 100~keV does not originate in localised regions, like the bright
X-ray knots and filaments or the primary blast wave, but is spread over the
whole remnant with a rather flat hardness ratio of the 8$-$10 and 10$-$15~keV
energy bands.  This result does not support an interpretation of the hard
X-radiation as synchrotron emission produced in the primary shock, in which case
a limb brightened shell of hard X-ray emission close to the primary shock front
is expected.  In fact a weak rim of emission near the primary shock front is
discernable in the hardest X-ray image but it contains only a few percent of the
hard X-ray emissivity.  The equivalent width of the Fe-K line blend varies by
more than an order of magnitude over the remnant, it is hard to explain this as
Fe-emission from the reverse shock heated ejecta given the ejecta temperature
and the age of the remnant.  The uniquely high wavelength-dispersive
RGS-spectrometer has allowed, for the first time, to extract monochromatic
images in several highly ionised element species with high spectral resolution.
We present here a preliminary result on the measurement of the \ion{O}{viii}
Ly-$\alpha$ and Ly-$\beta$ brightness distribution and brightness ratios.  The
large observed decrease of the Ly-$\alpha$/Ly-$\beta$ ratio going from the N
to the SE can be explained by small-scale (10~\arcsec) variations in the
$N_{\rm H}$ column over the remnant and the potential presence of resonance
scattering of the \ion{O}{viii} Ly-$\alpha$ photons in the limb brightened
shell.
 }

\maketitle

\section{Introduction}

The young galactic supernova remnant (SNR) Cassiopeia A (Cas A) is widely
believed to be the result of the core collapse of a massive star, probably an
early type Wolf-Rayet star (Fesen et al.  \cite{fesen}).  Cas A is classified as
an oxygen rich remnant since optical spectroscopic observations (Chevalier \&
Kirshner \cite{chevalier}) show the supernova ejecta (in the form of fast moving
knots) to contain mostly oxygen and oxygen burning products such as sulphur,
argon and calcium.  At all wavelengths Cas A has the appearance of a broken
shell with a radius varying between 1.6\arcmin\ to 2.5\arcmin.

One of the outstanding problems in the study of Cas A is the origin of a
recently discovered hard X-ray continuum in the spectrum of this remnant.  From
a theoretical point of view both synchrotron emission from shock accelerated
electrons and non-thermal Bremsstrahlung from electrons accelerated from the
tail of the thermal emission could be possible explanations for the hard
continuum.  In addition, high resolution spatially resolved spectroscopy of the
remnant in the X-ray domain is a powerful tool to study in detail the
distribution and physical properties of the (reverse) shock heated plasma.  The
bandwidth (up to 15 keV) of XMM-Newton and the presence of high
wavelength-dispersive spectrometers (RGS) offers an unique opportunity to
address these questions.  In this respect XMM-Newton strongly complements
Chandra, 
and this paper gives a first highlight of this capability.

\section{Observations}

A description of the instrument is given by Jansen et al.  (\cite{jansen}).  The
data were obtained in July 2000 with a net exposure time of 30~ks.  The
telescope was pointed at the centre of the remnant ($\alpha=23^{\rm h}23^{\rm
m}25^{\rm s}$ $\delta$=58\degr48\arcmin20\arcsec) and the telescope roll angle was
such that the RGS dispersion axis was aligned at 45\degr\ (NE) on the sky.

The raw data were processed with the development version of the XMM-Newton
Science Analysis System(SAS).  The RGS spectra were extracted by applying
spatial filters to the spectral image and the spectral order ($m=-1$) is selected
applying the appropriate pulseheight intervals to the CCD spectral camera.

\section{The hard X-ray data}

A high-energy tail has been observed in the X-ray spectrum of the supernova
remnant Cassiopeia A by the Compton Gamma Ray Observatory (The et al.
\cite{the}), BeppoSAX (Favata et al.  \cite{favata}) and the Rossi X-ray Timing
Explorer (Allen et al.  \cite{allen}).  Previous hard X-ray imaging
observations, for example using BeppoSAX (Vink et al.  \cite{vink}) indicated
that the hard continuum radiation originated predominantly in the W region of
the remnant, however this result was based on deconvolved images from a
typically 1\arcmin\ resolution telescope with moderate effective area.  The
combination of large collecting area in the energy band 4.0 to 12.0 keV and
angular resolution of a few arcseconds of XMM-Newton provides us with a unique
opportunity to search for the distribution and origin of this hard ``tail''.

\subsection{The X-ray spectrum above 4.0 keV}

\begin{figure}[!htb]
\includegraphics{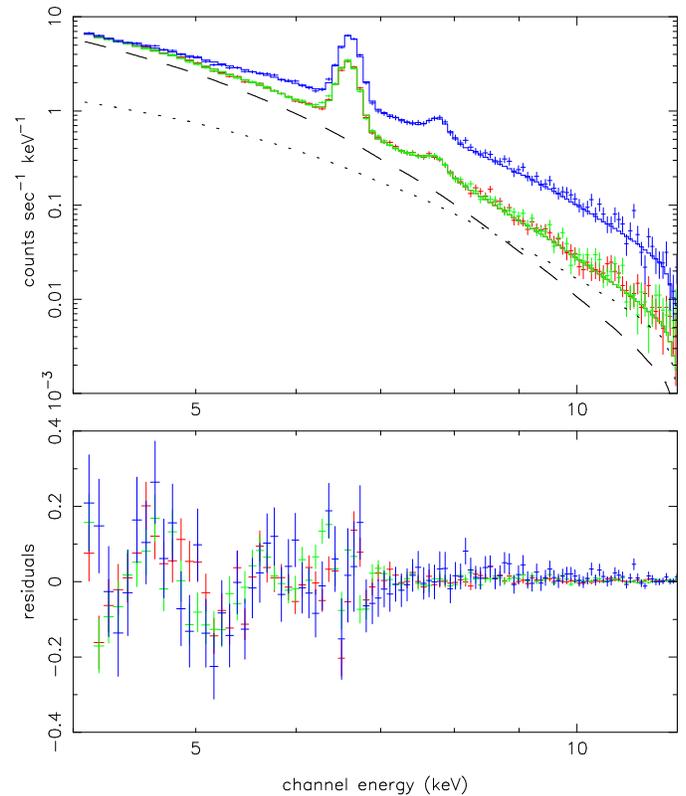}
\caption{Joint spectral fit of the hard tail from the complete remnant,
MOS 1 red, MOS 2 green and PN blue. The dashed and dotted line
indicate the Bremsstrahlung and power law components contributing
to the MOS continuum.}
\label{fig1}
\end{figure}

Below 4.0 keV the observed spectrum is a complicated combination
of spectral lines, continuum emission and the effects of interstellar
absorption but above 4.0 keV it is dominated by a smooth continuum
and the Fe-K emission line complex.
A joint spectral fit of the EPIC MOS and PN spectra was performed
in the energy band 4.0$-$12.0 keV for the complete remnant.
The model included a single Bremsstrahlung component plus
a power law continuum with photon index fixed at the previously
observed value (Allen et al. \cite{allen}) of 1.8 below a break energy
of 16~keV.
Gaussian line features were used to fit the obvious Fe and Ni line blends.
The resulting fit is shown in Fig.~\ref{fig1}.
The best fit Bremsstrahlung temperature is $2.56\pm 0.05$ keV with emission
measure $1.99\pm 0.04\times 10^{59}$~cm$^{-3}$ for a distance of 3.4~kpc and the
power law flux at 1 keV is $0.0416\pm 0.003$ photons cm$^{-2}$ s$^{-1}$
keV$^{-1}$.  The Fe line energy is $6.603\pm 0.001$ keV in the MOS and
$6.623\pm 0.002$ keV in the PN.  The errors quoted are statistical only and
indicate the high quality of the data, the difference between the MOS and PN
values may be attributed to systematic uncertainties which are at present of the
order of 10~eV.  We estimated the effect of potential pile-up in the PN camera
to be not larger than $\sim$3~\% for the brightest regions.  Fig.~\ref{fig1}
indeed shows that the MOS spectra, which do not suffer from pile-up, and the PN
spectrum are fully compliant.  The equivalent width of the Fe line is 1.02 keV
for the MOS and 1.11 keV for the PN.  The Ni line energy is $7.740\pm 0.015$ keV
in the MOS and $7.783\pm 0.015$ keV in the PN corresponding to \ion{Ni}{xxvi}
7.687 keV and/or \ion{Ni}{xxvii} 7.799 keV.  The equivalent width of the Ni line
was 0.130 keV for the MOS and 0.179 keV for the PN.  Allen et al.
(\cite{allen}) fitted the composite spectrum of Cas~A over a much broader energy
range extending out to 100~keV using two Raymond-Smith collisional ionisation
equilibrium (CIE) thermal components with k$T_1$=0.6 and k$T_2$=2.9~keV to fit
the soft band below 6~keV and a broken power law ($E_b$=15.9~keV) with photon
index 1.8 below $E_b$ and 3.0 above $E_b$ with a flux of
0.038~photons~cm$^{-2}$~s$^{-1}$~keV$^{-1}$ at 1~keV.  Favata et al.
(\cite{favata}), from the BeppoSAX observations, obtained a continuum
parameterized by two non-equilibrium ionisation (NEI) components with
k$T_1$=1.25~keV, k$T_2$=3.8~keV and a power law with photon index 2.95 with a
normalisation of 0.74 photons~cm$^{-2}$~s$^{-1}$~keV$^{-1}$ at 1~keV.  It is
most important to note that, irrespective of the exact parameterization of these
fits, they all share the fact that the X-ray flux above 8~keV becomes dominated
by the power law component.  More quantitatively, our spectral fit predicts a
ratio between the power law and the Bremsstrahlung component in the MOS cameras
of 0.29 in the interval 4$-$6~keV, 1.04 in the 8$-$10~keV band and 2.00 in the
10$-$12~keV hardest band.  We shall come back to this point in the next section
on the X-ray images.

\subsection{X-ray images above 4.0 keV}

\begin{figure}[!htb]
\includegraphics{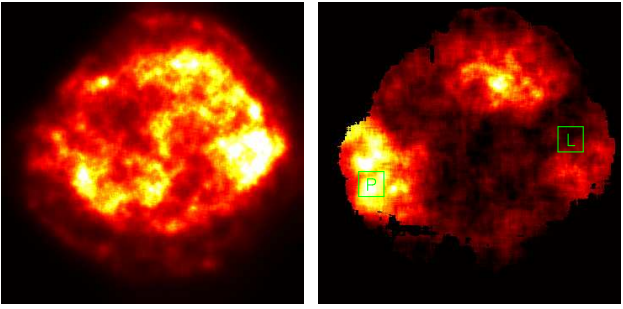}
\caption{Continuum 4.06$-$6.07 keV (left) and Fe K equivalent width (right)
black 0 keV to white 4 keV and above.}
\label{fig2}
\end{figure}

\begin{figure}[!htb]
\includegraphics{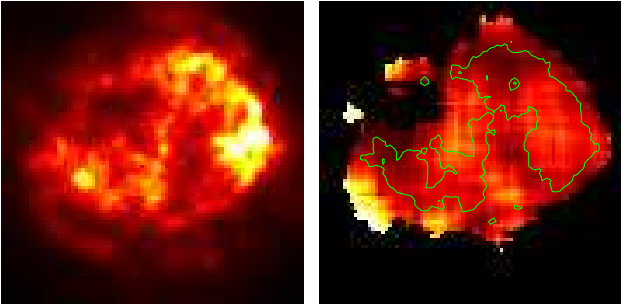}
\caption{Continuum 8.10$-$15.0 keV (left) and hardness ratio
(10.0$-$15.0 MOS+PN)/(8.10$-$10.0 MOS+PN) 0.15 black to 0.45 white.}
\label{fig3}
\end{figure}

The MOS and PN images are practically identical.
Fig.~\ref{fig2} shows MOS images
and Fig.~\ref{fig3} shows combined MOS and PN images.
The lefthand panel of Fig.~\ref{fig2} is the continuum map for 4.07$-$6.07 keV an
energy band in which no bright spectral lines are visible in either the EPIC MOS
or PN spectra.  The peak of the hard continuum occurs in the large knot to the
W but there is also extended emission throughout the volume of the remnant.
The outer rim (presumably just behind the blast wave) is visible but is very
faint compared with the central volume.

The righthand image of Fig.~\ref{fig2} is the Fe K line equivalent width derived
from the narrow energy band 6.20$-$6.92 keV.  The continuum used to estimate the
equivalent width was estimated by interpolation from the bands 4.06$-$6.07 keV
and 8.10$-$15.0 keV.  The Fe line is very bright relative to the continuum on
the outer extremities of the SE knots, i.e.  region marked P, but is rather
weaker in the W where the continuum is bright, i.e.  region marked L.
The lefthand panel of Fig.~\ref{fig3} is the hard continuum above the
Fe K line, 8.10$-$12.0 keV, which is very similar to the softer continuum
image in Fig.~\ref{fig2}. In order to try and locate where the
hard X-ray tail originates we calculated the hardness ratio
(10.0$-$15.0)/(8.10$-$10.0) using both the MOS and PN cameras to give
the best possible statistics. The result is shown in the righthand panel
of Fig.~\ref{fig3}. The contour superimposed in green is taken
from the hard continuum map
8.10$-$15.0 keV and contains 37~\% of the total counts in this energy band,
the hardest region in the W contains less than 3~\% of the total.
The remaining 63~\% of the hard flux is spread out over the rest of
the remnant. There are significant changes in hardness over the remnant
but all the bright features within the green contour
have a remarkably similar spectrum above 8.0~keV.

From the spectral fits described in the previous section, it is clear that the
photon flux observed in the continuum image in Fig.~\ref{fig3} is dominated by
the hard tail (power law) component.  XMM-Newton clearly detects the hard X-ray
tail from the remnant but the hard X-ray image and the hardness ratio indicate
that this flux does not predominate in a few localised regions, but pervades the
whole remnant in a distribution similar to the softer thermal components.

\subsection{The Fe K emission}

Simpler spectral fits using a single Bremsstrahlung and one gaussian line were
used to quantify the variation of the continuum and Fe K emission over the
remnant.  Region L in Fig.~\ref{fig2} corresponds to the brightest hard
continuum and a rather low Fe equivalent width and region P to rather weak
continuum with the highest Fe K equivalent width.  The same model was also used
to fit the composite spectrum.  The results are summarised in
Table~\ref{specfit}.  The apparent temperature in these fits is a measure of
hardness and is higher than the value reported above because the power law
continuum has been omitted for simplicity.  It is clear that the temperature for
the composite fit is similar to that of the small regions (within the
statistical accuracy), again indicating that the hard flux is distributed
throughout the remnant.
\begin{table}
\caption[]{Results of joint spectral fits using a single Bremsstrahlung
and gaussian line 4.0$-$12.0 keV. Listed are the temperature k$T$,
line centroid $E_\ell$, equivalent width $EW$ 
and line sigma width $\sigma$. The first values for $E_\ell$, $EW$
and $\sigma$ are for the MOS cameras and the second from the PN.}
\begin{tabular}{|l|ccc|}
\hline
Parameter    & all & L & P \\
\hline
k$T$ (keV)     & 3.53$\pm$0.02 & 3.41$\pm$0.14 & 3.15$\pm$0.34 \\
$E_\ell$ (keV) & 6.60       & 6.60       & 6.67 \\  
               & 6.62       & 6.61       & 6.68 \\
$EW$ (keV)     & 0.95$\pm$0.02 &  0.57$\pm$0.08 & 6.19$\pm$0.56 \\
               & 1.01$\pm$0.02 &  0.43$\pm$0.18 & 5.28$\pm$0.53 \\
$\sigma$ (keV) & 0.083 &  0.079 &  0.047 \\ 
               & 0.078 &  0.042 &  0.031 \\             
\hline
\end{tabular}
\label{specfit}
\end{table}
As shown in Fig.~\ref{fig2} and quantified in Table~\ref{specfit}
the variation in equivalent width of the
iron line is very large. We attempted to fit the
composite spectrum from MOS 1 + 2 + PN 4.07$-$12.0 keV
with a simple NEI model (see for example Borkowski \cite{bork})
plus a power law.  The power law was again fixed at index 1.8 and
the temperature k$T$ fixed at 2.6 keV.
This gave a good fit to the continuum with the Fe line blanked out.

A reasonable fit was obtained for an ionisation timescale parameter of
$5.0\times 10^{11}$~s~cm$^{-3}$ or greater.  For an age of 340 years, ($10^{10}$
seconds) this implies a density of $>50$ cm$^{-3}$ which is rather large.  This
fit gave about the correct equivalent width for the Fe line (assuming solar
abundances), but did not fit the line profile very well, presumably because the
line is Doppler shifted and/or broadened.  It also gave a reasonable fit to the
Ni line feature at around 7.7 keV.  Lower more reasonable values of the time
scale $5.0\times 10^{10}$~s~cm$^{-3}$ corresponding to a density of 5~cm$^{-3}$
predict a much larger equivalent width (a factor of 4 too large), a line energy
which is too low and no Ni line at 7.7~keV.

Using the same model on region P in the SE where the equivalent width
of Fe is very high gave a poor fit. No combination of temperature or
time scale could produce a strong enough line at the correct energy.
The E knot is clearly anomalous with abnormally
high Fe abundance and/or extreme non-thermal equilibrium ionisation.

\subsection{Discussion}

The spectral form of the non-thermal high-energy ``tail'' is not inconsistent
with a simple model of synchrotron emission from SNRs Reynolds \cite{reynolds}.
However in such a model the electrons need to be accelerated to energies of tens
of TeV at the primary shock and the associated synchrotron emission would be
expected to be concentrated in the compressed magnetic field just inside the
shock front.  The hard X-ray continuum maps from XMM-Newton indicates that the
8.0$-$15.0~keV flux, predominantly due to the previously reported high-energy
tail, does not originate from a few localized regions such as X-ray bright knots
and filaments, nor does it originate from a limb brightened (fractionary) shell
close to the shock front generated by the primary blast wave.  In fact a low
brightness outer ring structure, presumably associated with the primary shock,
can be discerned in the hard X-ray image but it contains only a few percent of
the total hard X-ray flux.  Therefore the hard X-ray image is morphologically
inconsistent with the simple synchrotron model developed by Reynolds
(\cite{reynolds}).  An alternative explanation for the observation of hard X-ray
tails in the spectra of supernova remnants is the presence of non-thermal
Bremsstrahlung generated by a population of suprathermal electrons (Asvarov et
al.  \cite{asvarov}; Laming \cite{laming}).  One might expect in this case some
degree of correlation with the strong line emitting regions, however abundance
variations make this potential correlation ambiguous.  A search of Cas~A for
regions with spectra devoid of line emission, as a possible tracer of
synchrotron emission, was unsuccesful although Chandra observations indicated
the presence of such a region just beyond the westernmost tip of the N
rim of the remnant (Hughes et al.  \cite{hughes}).  We specifically checked this
region and still confirm the presence of, relatively weak, line emission, with
for example equivalent widths of 200$-$300~eV for the He-like triplet of Si, S
and Fe.  We cannot exclude some contamination by line emission from neighbouring
regions due to the wings of the XMM psf, on the other hand the Chandra data for
this region also show line residuals above the continuum fit presented by Hughes
et al.  In summary we have not spotted any positive evidence, either
morphologically or spectrally, for the presence of synchrotron emission in Cas~A
and, by implication, 
the associated TeV electrons.  Consequently, whether the hard X-ray tail is
thermal or non-thermal remains an open question.

\section{High-resolution spectroscopic data}
 
{\subsection{Analysis \& results}
\begin{figure}
\resizebox{\hsize}{!}{\includegraphics[angle=-90]{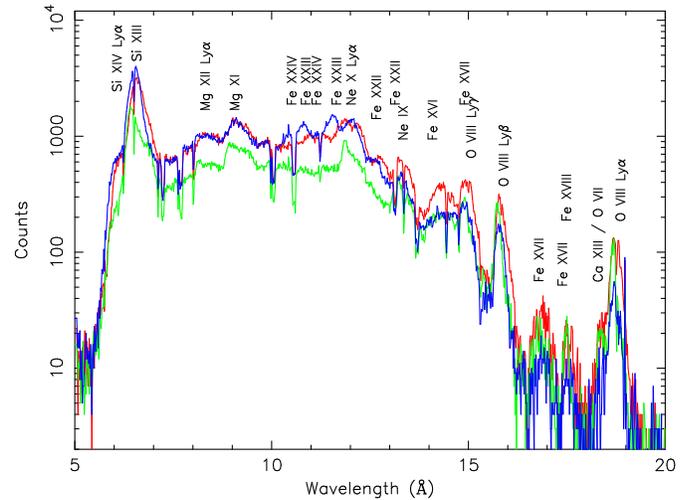}}
\caption[]{First order ($m=-1$) RGS spectra of Cas~A for the three regions. Line 
identifications of the principle lines are provided.}
\label{fig:spectra}
\end{figure}
Fig.~\ref{fig:spectra} displays spectra from three
extraction regions over the
remnant situated in the N, NE and SE of the remnant.  The data from
RGS1 and RGS2 have been combined in order to increase the statistical
weight of the observation.  Several lines of highly ionized species
of Si, Mg, Ne, Fe L and O are
detected in the spectrum.  Due to interstellar absorption no features can be
measured long-ward of $\sim$20~\AA.  The analysis
and interpretation of these spectra
will be the subject of a forthcoming paper.

As a first result for this paper we extracted images of the \ion{O}{viii}
Ly-$\alpha$ and Ly-$\beta$ lines
to probe small scale variations in absorption effects over this
part of the remnant and to investigate the potential presence of
resonance scattering in the limb brightened
shells viewed edge-on. The temperature range relevant for Cas~A
has no influence on the Ly-$\alpha$/Ly-$\beta$ ratio.
Since RGS is a slitless spectrometer it is
possible to extract dispersed monochromatic images of Cas~A.  These images were
converted from wavelength to spatial coordinates using the equation
$\Delta\lambda = d\sin(\alpha)\Delta\phi$, where $\Delta\phi$ is the offset
along the dispersion direction, $\Delta\lambda$  the wavelength shift,
$\alpha$ the angle of incidence on the gratings and $d$ the line distance
of the gratings.  However, any Doppler broadening is also convolved along the
dispersion direction and, depending on its magnitude, Doppler broadening could
distort the RGS images.
\begin{figure}
\resizebox{\hsize}{!}{\includegraphics{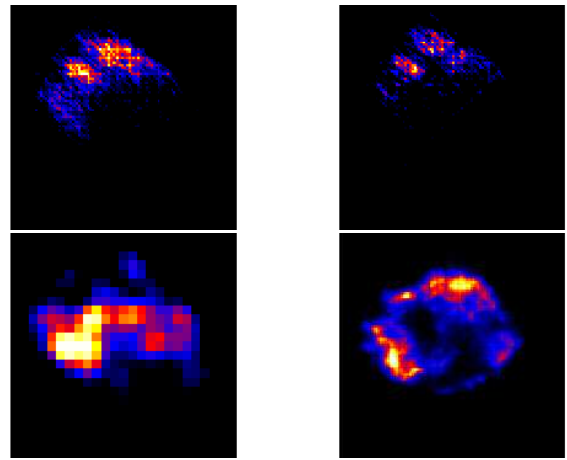}}
\caption[]{RGS images of \ion{O}{viii} Ly-$\beta$ (top left)
and Ly-$\alpha$ (top right), the dispersion direction
runs from top-left to lower-right at 45 degrees with the vertical.
Emission is seen in three blobs 
in the E and N rim. The ratio map of the Ly-$\beta$ to
Ly-$\alpha$ images is shown at the bottom left panel. 
The enhanced spot on the SE rim of the image indicates
a Ly-$\alpha$ emission deficit in this region. For comparison
we show a soft band MOS image (0.5$-$1.2~keV) in the lower right panel.}
\label{fig:oimages}
\end{figure}
The \ion{O}{viii} Ly-$\alpha$ and Ly-$\beta$ maps are presented in
Fig.~\ref{fig:oimages}.  The oxygen emission is seen to originate from three
blobs on the E and N rim of the remnant.  
The emission is faintest in the
SE blob where there also seems to be a Ly-$\alpha$ deficit compared to
Ly-$\beta$.  To investigate this we produced cross-dispersion image
profiles of the Ly-$\alpha$ and Ly-$\beta$ lines by integrating over the
dispersion direction.  The cross-dispersion profiles together with the
Ly-$\alpha$ to Ly-$\beta$ ratio have been
plotted in Fig.~\ref{fig:xdsp} (note that this is the inverse of the
ratio plotted in the brightness distribution in Fig.~\ref{fig:oimages} because
of the low Ly-$\alpha$/Ly-$\beta$ ratio).  The ratio
plot seems to steepen gradually from a value of
$\sim$0.5 to $\sim$0.15$-$0.20, 
at a cross-dispersion above +1\arcmin.
This decrease seems to indicate a Ly-$\alpha$ emission
deficit in this region.  
The same effect is seen in the ratio image presented in
Fig.~\ref{fig:xdsp}.
\begin{figure}
\resizebox{\hsize}{!}{\includegraphics{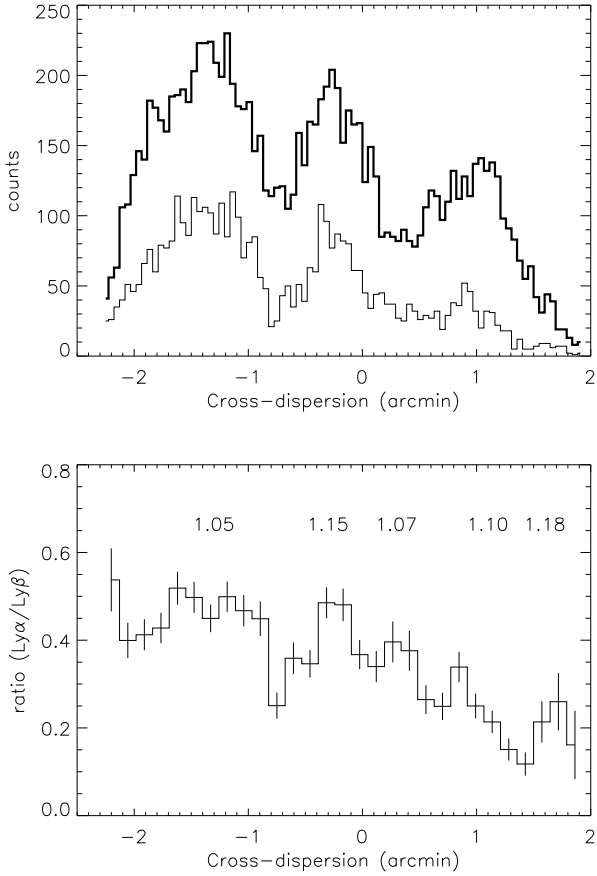}}
\caption[]{Upper panel: cross dispersion profiles of \ion{O}{viii}
Ly-$\beta$ (thick line) and  \ion{O}{viii} Ly-$\alpha$
(thin line). Lower panel:
ratio of Ly-$\alpha$ to Ly-$\beta$. Numbers indicate
estimated $N_{\rm H}$ values from Keohane (1998), in units of
$10^{22}$~cm$^{-2}$.}
\label{fig:xdsp}
\end{figure}

\subsection{Discussion}

Three possible causes can be identified for explaining the \ion{O}{viii}
Ly-$\alpha$ deficit in the SE part of Cas~A, i.e.
line blending, $N_{\rm H}$ variation and resonance scattering.

(1) A line blending effect due to the reduced spectral resolution arising from
the finite source extent causes the Ly-$\beta$ line to be "polluted"
by neighbouring lines from the Fe-L complex. Given the limited
angular extent of the bright emission region in the dispersion direction, i.e.
$<$1\arcmin, and the fact that Cas~A is an oxygen-rich remnant, this effect
can be neglected, i.e. the oxygen emission lines strongly dominate the
Fe-L lines and the spectral resolving power is still adequate (R$\sim$130). 

(2) The potential influence of variations in the column density $N_{\rm H}$
in this region can be assessed by using the column densities given by
Keohane (\cite{keohane}). These data take into account both the \ion{H}{i}
and molecular absorption and have an average resolution of 30\arcsec.
In the E part of Cas~A they centre at $1.10\times 10^{22}$~cm$^{-2}$ with a 
full-width spread ranging from $1.00-1.20\times 10^{22}$~cm$^{-2}$ (only the
bright spot Cas~A W shows a larger column of about 
$1.4\times 10^{22}$~cm$^{-2}$).
Applying this full range going from N (least absorbed)
to the SE bright rim
(most absorbed) yields a decrease of the \ion{O}{viii} Ly-$\alpha$/Ly-$\beta$
ratio of 1.4. The observed decrease of the Ly-$\alpha$/Ly-$\beta$
ratio, displayed in Fig.~\ref{fig:xdsp}, appears to be substantially larger, i.e.
a factor $\ge$3 from about 0.45 to 0.15. Smaller-scale
variations could be caused
by small scale structure in the $N_{\rm H}$ value on a $\sim$10\arcsec\
angular scale.
To fully explain the decrease observed a column density variation of
$6\times 10^{21}$~cm$^{-2}$ (i.e. from $1.0-1.6\times 10^{22}$~cm$^{-2}$) would have to
be present, which  seems rather unlikely.

(3) Alternatively a decrease in the 
Ly-$\alpha$/Ly-$\beta$ ratio could also be introduced by the presence of
resonance scattering of the \ion{O}{viii} Ly-$\alpha$ photons
in the X-ray bright rims if viewed edge-on (Kaastra \& Mewe \cite{kaastra}).
A factor of 2$-$3 reduction
in Ly-$\alpha$ intensity can be explained by an optical depth in
the Ly-$\alpha$ line of 2 or more (following the discussion in
Kaastra \& Mewe). For typical conditions in the rims
of Cas~A, an optical depth larger than 2
is expected to be present if the micro turbulent velocity 
is smaller than $\sim$700~km/s.

\begin{acknowledgements}
The results presented are based on observations obtained with XMM-Newton,
an ESA science
mission with instruments and contributions directly funded by
ESA Member States and the USA.
We thank the referee J. Ballet for his helpful comments and suggestions.
\end{acknowledgements}

\end{document}